\documentclass[a4paper,twocolumn]{article}
\usepackage{graphicx}
\usepackage{epstopdf}
\usepackage{float}
\usepackage{latexsym}
\usepackage{multicol}
\usepackage{mathrsfs}
\usepackage{amsmath}
\usepackage{amssymb}
\usepackage{listings}
\usepackage[super,sort&compress,comma]{natbib}
\usepackage[utf8]{inputenc}
\usepackage{url}
\usepackage{sidecap}
\usepackage{dsfont}
\usepackage[dvipsnames,svgnames]{xcolor}
\usepackage{siunitx}
\usepackage{subfigure}
\usepackage[export]{adjustbox}
\usepackage[format=plain,justification=justified,singlelinecheck=false,labelfont=bf,labelsep=space]{caption}
\usepackage[left=1.5cm, right=1.5cm, top=1.785cm, bottom=2.0cm]{geometry}
\usepackage[compact]{titlesec}
\usepackage{balance}

\titlespacing*{\section}{0pt}{4pt}{4pt}
\titlespacing*{\subsection}{0pt}{15pt}{1pt}

\newcommand{\blue}[1]{\textcolor{black}{#1}}

\begin{document}

\twocolumn[
  \begin{@twocolumnfalse}
  
\centering

\begin{minipage}{0.8\linewidth}

\title{The influence of surface roughness on the adhesive interactions and phase behavior of suspensions of calcite nanoparticles}
\author{\normalsize Juan D. Olarte-Plata$^a$, G{\o}ran Brekke-Svaland$^a$ and Fernando Bresme$^{a}$}
\date{}        
\maketitle
We investigate the impact of nanoparticle roughness on the phase behaviour of suspensions \blue{in models} of calcium carbonate nanoparticles. 
We use a Derjaguin approach that incorporates roughness effects and interactions between the nanoparticles modelled with a combination of DLVO forces and \blue{hydration} forces, derived using experimental data and atomistic molecular dynamics simulations, respectively. Roughness effects, such as atomic steps or terraces appearing in mineral surfaces result in very different effective inter-nanoparticle potentials.
Using stochastic Langevin \blue{Dynamics} computer simulations and the effective interparticle interactions we demonstrate that relatively small changes in the roughness of the particles modify significantly the stability  of the suspensions. \blue{We propose that the sensitivity of the phase behavior to the roughness is connected to the short length scale of the adhesive attraction arising from the ordering of water layers confined between calcite surfaces}. Particles with smooth surfaces feature strong adhesive forces, and form gel fractal structures, while small surface roughness, of the order of atomic steps in mineral faces, stabilize the suspension. We believe that our work helps to rationalize the contrasting experimental results that have been obtained recently using nanoparticles or extended surfaces, which provide support for the existence of adhesive or repulsive interactions, respectively. We further use our model to analyze the synergistic effects of roughness, pH and ion concentration on the phase behavior of suspensions,  connecting with recent experiments using calcium carbonate nanoparticles.
\vspace{1cm}

\end{minipage}

\end{@twocolumnfalse}
]

\let\thefootnote\relax\footnote{\textit{$^{a}$~Department of Chemistry, Imperial College London, Molecular Sciences Research Hub, White City Campus,
80 Wood Lane, London W12 0BZ, UK. E-mail: j.olarte@imperial.ac.uk, g.svaland15@imperial.ac.uk, f.bresme@imperial.ac.uk}}

\section{Introduction}

Nanoparticle suspensions are widely used in soft materials; food stuffs, pharmaceuticals, and high performance nanofluids 
\cite{Saidur2011}. Suspensions of calcite nanoparticles (CN)
are widely employed to manufacture cements for the 
construction industry~\cite{Coussot2007,Dalas2015} and calcite powders, with nanoparticles diameters between 100's of \si{nm} to \si{\micro\metre}, are used to make pastes \cite{Liberto2017}, for the manufacture of paper as well as building materials. CN have been used recently to functionalize living cells~\cite{cncells}.  The aggregation of CN is also relevant in self-assembly processes related to biomineralization~\cite{C0NR00697A}. NP interactions 
determine the phase behavior and rheology of the suspension, and therefore understanding these interactions is important both in industrial applications and assembly processes occurring in Nature. 

Calcite surfaces are prone to dissolution and crystallization, as a consequence the surfaces can develop different levels of roughness. 
Experimental studies of extended calcium carbonate surfaces using the surface forces apparatus \cite{Dziadkowiec2018}, indicate that roughness effects can play a significant role in determining the interactions between calcite surfaces. Roughness effects are manifested in the measurement of repulsive interactions when the surfaces are immersed in water. Experimental studies of CN aqueous suspensions, with nanoparticle diameters of 60-70 \si{nm}, demonstrated the formation of gel phases \cite{Liberto2017}, and therefore the existence of strong adhesive interparticle interactions
\cite{Liberto2019}.  The mechanical strengthening of calcium carbonate pastes obtained from recrystallization of amorphous calcium carbonate and vaterite mixtures, has been rationalized considering the smoothing of the nanoparticle surfaces at grain contacts \cite{Rodriguez2018}. Recent molecular dynamics simulations of flat calcium carbonate surfaces immersed in water do also provide support for attractive interactions. Strong adhesive minima were observed at inter-surface separations of about 1 nm. This strong adhesion is mostly driven by the layers of water molecules adsorbed at the calcite surfaces. Shifts in the relative position of the surface planes was shown to influence the interaction strength too~\cite{Svaland2018}.
While the simulation results and the behaviour of nanoparticle suspensions seem consistent with each other regarding the observation of adhesive interactions, 
the experiments on extended surfaces demonstrated on the other hand the existence of 
repulsive interactions. We believe that the consideration of surface roughness might help to understand these results.

Understanding the role of nanoparticle roughness on the phase behavior of a suspension 
requires an extension of the existing theoretical models, such as the DLVO theory \cite{Israelachvili2011}, to incorporate roughness effects. The latter 
effects have indeed been considered in previous works. %
Surface and chemical heterogeneities were identified as potential contributors to the discrepancies between theory and experiments \cite{Kihira1992, Elimelech1995, Walz1998}, prompting the implementation of surface roughness in theoretical models~\cite{Czarnecki1980,Duval2004,Bhattacharjee1998,Hoek2006,Huang2010,Yang2011,Parsons2014a}.  These works demonstrated that the 
roughness can lead to interactions that differ from the DLVO potential. 
The double layer repulsion arising from the calcite surface charge is expected to be much smaller than the solvation forces at nanometer inter-surface separations, given the low surface potential of calcite (-0.02 C/m${^2}$ \cite{Lee2016}), and therefore should contribute little at this range of distances. Previous studies have highlighted the impact of surface roughness on the local surface charge, such as steps or terraces, leading to a ten-fold increase with respect to flat calcite surfaces \cite{Wolthers2012}. Particle size is also known to modify the surface charge in titanium dioxide nanoparticles \cite{Holmberg2013}. However, experiments on calcium carbonate nanoparticles have reported the formation of gel structures \cite{Liberto2017,Liberto2019}, which support the existence of adhesive interactions. This suggest that the double layer repulsion arising from surface charges may not be dominant at very short intersurface distances.

We adopt in this work a convolution approach to investigate the influence of nanoparticle roughness on the particle interactions. The approach builds on the model developed by Parsons \textit{et al.}~\cite{Parsons2014a}, whereby the force between flat surfaces is convoluted with a probability distribution that defines the roughness of the surface of interest. A key input for this approach is the solvent mediated interactions between flat calcite surfaces, which were computed recently using state of the art forcefields and molecular simulations. We use the resulting effective potentials to investigate the phase behavior of colloidal suspensions by means of Langevin Dynamics computer simulations. 
The surface roughness is shown to have 
a strong impact on the 
adhesive minimum found in flat surfaces, which 
disappears in rough colloids, even for small RMSD roughness of the surface $\sim 2.0$~\AA, rendering 
repulsive interactions dominant. 
We investigate the interplay of
adhesion, DLVO forces and roughness effects in defining the interactions between CN, as a function of the ionic strength of the aqueous solution.

\begin{figure}[t]
\centering
\includegraphics[width=0.9\linewidth]{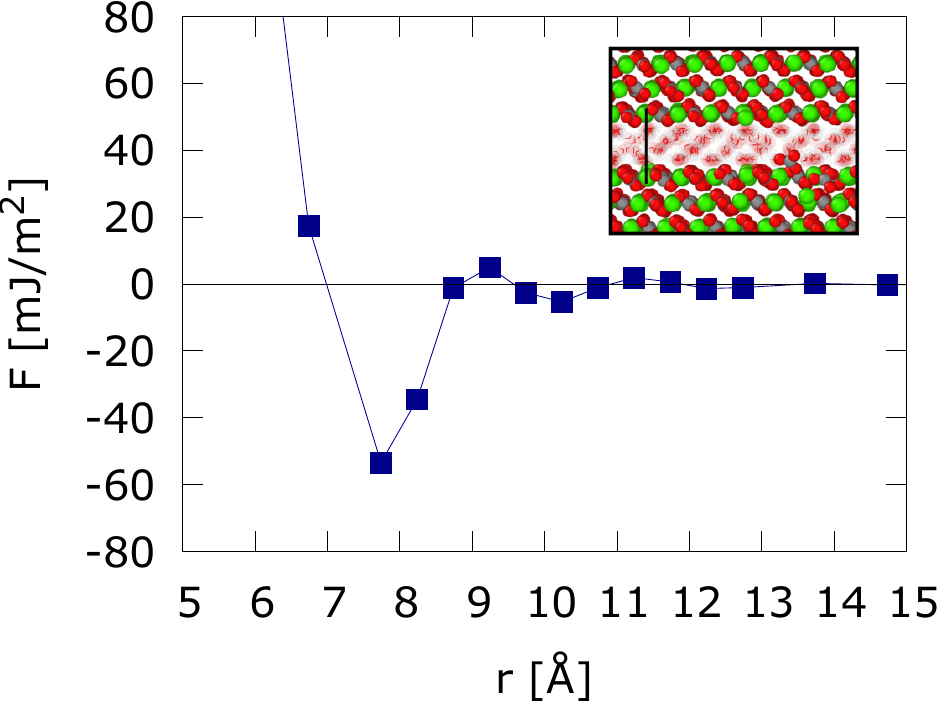} \\
\captionof{figure}{\footnotesize Free energy interaction  
between two atomically flat calcite $(10\bar{1}4)$ surfaces. The inset shows the structuring of water confined between two calcium carbonate ($10\bar{1}4$) surfaces in registry at $D=$~7.5\,\AA \, intersurface separation. Results taken from reference \cite{Svaland2018}.}
\label{fig:epitaxy}
\end{figure}

\section{Methods}

\subsection{Free energies of flat surfaces}

To model the interactions between the CN surfaces we used the DLVO theory and the structural forces computed in reference 
\cite{Svaland2018}. The free energy curve was obtained by computing the interaction forces at different inter-surface distances between $r_0$ and $r$. The corresponding change in free energy in this distance interval is given by, $\Delta F(r) = F(r)-F(r_0) = -\int _{r_0}^{r} f(s)ds$, where $f(s)$ is the total force acting on the surfaces as a function of the inter-surface separation, $r$. $r_0$ is a reference distance defining the zero of the free energy at long separation. The structural forces obtained in ref. \cite{Svaland2018} quantify the solvation forces arising from strong water layering induced by the confining surfaces.

We 
reproduce in Fig.~\ref{fig:epitaxy}, the solvent mediated interaction free energy of two flat calcite surfaces corresponding to the ($10\bar{1}4$) plane. The 
free energy features a strong adhesive minimum at $r<1$~nm~\cite{Svaland2018}. \blue{We note that surface hydration can result in long range repulsion. Such effect has been observed in soft interfaces too~\cite{Faraudo2005}.
However, the calcite $(10\bar{1}4)$ surface induces the ordering of the interfacial water molecules, which adopt a checkerboard structure. This ordering has been observed both in experiments and computer simulations~\cite{STIPP19943023,songen2016,raiteri2010,Svaland2018}. When the two surfaces are put in close contact (nanometer distance) the ordering of the water layers spans the whole confined region, leading to strong adhesive force. Changes in the inter-surface distance disrupt the structure of the water layers, resulting in either narrow adhesive minima or repulsion at short distances (see r$<$ 7.5~\AA~in Figure 1). Strong adhesive minima have also been reported in mesoscopic models of Calcium Silicate Hydrate nanoparticles studies of cement matrices\cite{yu2015}.}

\blue{In our work we propose that the adhesive forces operate betwen calcite nanoparticles (several 10's of nm in diameter), where smooth surfaces might be present. A significant amount of hydrated ions at the surfaces could lead to rough surfaces and disrupt the adhesive force. However, previous studies concluded that ion-ion correlations do not play a significant role in the formation of calcite pastes~\cite{Liberto2019}. The estimated charge density is fairly low, $\sim$ 0.1 charges/nm$^2$. Furthermore, it has been reported that calcite nanoparticles feature smooth surfaces (see TEM images in ref.~\cite{Liberto2019}). The existence of such smooth surfaces in these small nanoparticles might well be compatible with the estimated, much longer $\sim 1 \mu$m, length scales of growth-induced roughening~\cite{bahareh2019}.  }

We demonstrate below that small 
RMS surface roughnesses $\rho\sim 2.0$~\AA, defined as the standard deviation of the surface height distribution with respect to the average surface height,
modify the attractive interaction, leading to strong hydration repulsion shor interparticle distances. Combining the roughened hydration interaction with the DLVO theory then yields effective potentials that can be used to investigate 
the phase behaviour of the colloidal suspensions.%}
We consider below experimental conditions corresponding to different ionic strengths and pH, which result in different double layer repulsions, which are modelled with the DLVO theory and the Derjaguin approximation.

\subsection{Incorporation of roughness contributions to the CN effective interactions} 

We include roughness effects over mesoscopic length scales by employing the Derjaguin approximation. As a hypothesis, we assume that the lateral size of rough domains is small, hence we use in this work a single RMS parameter to describe the topography of the surface. With regards to the magnitude of the roughness, experimental studies using AFM shown in Figs. 2A and 2B show the formation of atomic steps on the $(10\bar{1}4)$ surface \cite{Vavouraki2010}, with an RMS of the order of few angstroms, which we take  as an indicative value for the characteristic roughness of the nanoparticles. Adsorption of ions would introduce similar length scales due to disruption of hydration layers. The force, $F_{r}$, between two spherical particles is defined in terms of the inter-surface distance $r$. This force is determined by the interaction energy per unit area between two flat surfaces $W(r)$:

\begin{equation}
 F(r) = 2 \pi R_{eff} W(r)
 \label{eq:derjaguin}
\end{equation}

\noindent where $R_{eff}^{-1} = R_{1}^{-1} + R_{2}^{-1}$ is the effective radius of curvature, with $R_{1}$ and $R_{2}$ being the radii of the interacting spherical particles. The interaction energy between the two colloids can then be obtained from:

\begin{equation}
 U(r) = -\int \limits_{r_{0}}^{d} F(r) dr + U(r_{0})
 \label{eq:potential}
\end{equation}
\noindent where $U(r_{0})$ is an integration constant that defines the zero of energy for the potential at large inter-colloidal distances.

To calculate the 
roughness on the inter-particle interactions we introduce the height function, $f_{h}$, which is define by the 
Gaussian distribution:

\begin{equation}
 f_{h} = \dfrac{e^{-(r-h)^{2}/(2 \rho^{2})}}{\rho \sqrt{2 \pi}},
 \label{eq:roughness}
\end{equation}

\noindent where $r$ represents the surface-to-surface distance, $h$ represents the deviation from the reference surface $r$, and $\rho$ is the standard deviation that quantifies the degree of roughness of the surface. $\rho=0$ corresponds to a flat surface. Combining Eqn. (3) with the Derjaguin equations~(1) and (2), we obtain a ``roughened" Derjaguin approximation:

\begin{equation}
 F_{r}(r) = 2 \pi R_{eff} \int \limits_{-\infty}^{\infty} W(h) \dfrac{e^{-(r-h)^{2}/(2 \rho^{2})}}{\rho \sqrt{2 \pi}} dh
 \label{eq:roughderjaguin}
\end{equation}

\begin{figure*}[t]
\centering
\includegraphics[width=0.8\linewidth]{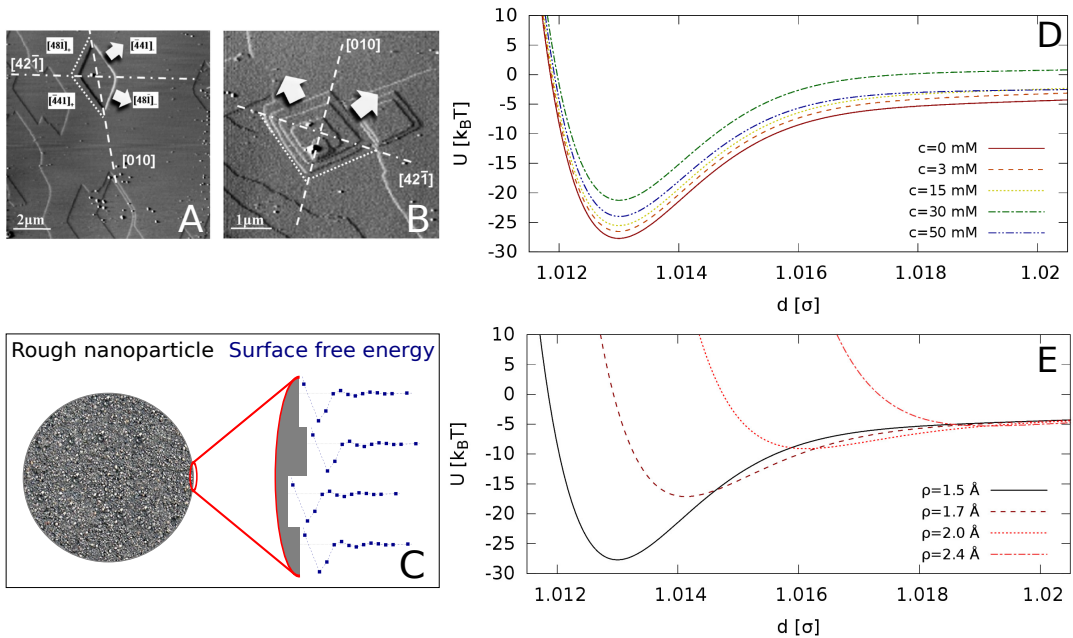}
\captionof{figure}{\footnotesize AFM images of a calcite $(10\bar{1}4)$ surface in A) deionized water and B) supersaturated solution with respect to calcite. The arrows indicate the crystallographic directions of terrace growth. Reprinted with permission from Vavouraki \textit{et. al.}, \textit{Crystal Growth and Design} 2010, 10, 1 60-69 \cite{Vavouraki2010}. Copyright 2010 American Chemical Society. C) Schematic representation of our model for a rough nanoparticle. The different surface heights result in an average inter-particle interaction given by the ``roughened'' Derjaguin approximation in Eq. (\ref{eq:roughderjaguin}), which averages the interaction of different surface heights (see detail of surface in zoomed image) by their probability distribution. D)  Inter-particle interactions for varying concentration of Ca(OH)$_2$, for surface roughness of $\rho=1.5$~\AA~(longer range included in the SI). E) Inter-particle interactions as a function of the particle roughness (increasing $\rho$ from bottom to top), for the zero ion conditions ($c=0$ mM).}
\label{fig:schematic}
\end{figure*}

This approach has been pioneered by Parsons \textit{et al.} to incorporate surface roughness in the theoretical calculation of surface forces \cite{Parsons2014a}. These authors showed that roughness amplifies the long range behaviour of DLVO forces, and shifts the repulsive branch detected in surface force measurements to a longer distance. The shift scales with the Root Mean Square roughness of the surfaces.

\subsection{DLVO interactions}
We model 
the double layer repulsion and van der Waals attraction between the calcium carbonate nanoparticles by means of the DLVO theory, implemented for the spherical geometry. The interaction energy as a function of the surface-to-surface distance, $r$, is given by \cite{Israelachvili2011}:

\begin{equation}
 U_{DLVO}(r) = \dfrac{-AR}{12r} + \dfrac{1}{2} R Z e^{-r/\lambda}
 \label{eq:dlvo}
\end{equation}

\noindent where $A$ is the Hamaker constant, $R$ is the particle radius, $\lambda$ is the Debye length and $Z$ is a pre-exponential factor given by:

\begin{eqnarray}
 Z & = & 64 \pi \epsilon_{0} \epsilon \left( \dfrac{k_{B}T}{e} \right)^2 \tanh^2 \left(\dfrac{e \zeta}{4 k_{B} T} \right) \nonumber \\
 & \approx & 4 \pi \epsilon_{0} \epsilon \zeta^2
 \label{eq:Zdlvo}
\end{eqnarray}

\noindent where $\epsilon_{0}$ is the vacuum permittivity, $\epsilon$ is the relative permittivity of water, $e$ is the elementary charge, and $\zeta$ is the Zeta potential. The second line of Eq. (\ref{eq:Zdlvo}) represents the Debye-H\"uckel approximation, valid for small surface potentials ($<25$ mV) \cite{Israelachvili2011}. The Debye length is related to the ionic strength, $I$, by:

\begin{equation}
 \lambda = \sqrt{\dfrac{\epsilon_0 \epsilon k_{B} T}{2 e^2 I}}
\end{equation}

To define the 
DLVO interactions we use information from the speciation analysis reported in reference \cite{Liberto2019}, which yields the Debye length and Zeta potential of the calcium carbonate surfaces as a function of the initial concentration of Ca(OH)$_2$. The values used in the present study are summarized in Table \ref{tab:conditions}.

\begin{table}[t]
\centering
\begin{tabular}{llll}
\hline
\textbf{c (mM)} & \textbf{I (mM)} & \textbf{$\lambda$ (nm)} & \textbf{$\zeta$ (mV)} \\ \hline
0               & 0.73            & 11.1                    & 9.3                   \\
3               & 0.47            & 13.8                    & 10.9                  \\
15              & 1.7             & 7.3                     & 12.5                  \\
30              & 10.2            & 3.0                     & 18.6                  \\
50              & 43.2            & 1.4                     & 18.6                  \\ \hline
\end{tabular}
\caption{\footnotesize Ionic strength (I), Debye length ($\lambda$) and Zeta ($\zeta$) potential \blue{of calcite nanoparticles of size 70 \si{nm}, as a function of the concentration of Ca(OH)$_2$. The data are taken from reference~\cite{Liberto2019}, and were obtained using a chemical speciation analysis and Zeta potential measurements.}}
\label{tab:conditions}
\end{table}

\subsection{Simulation methods}

We study a system of $N=1000$ nanoparticles dispersed in water using Langevin dynamics (LD), with water modelled
as a continuum. 
The particles were randomly inserted in a simulation box of  volume, $V$, at the chosen packing fraction of the suspension, given by $\phi = (\pi \sigma^{3} N)/(6 V)$. The interparticle interactions were defined using the approach discussed in sections 2.1-2.3. 

We used the viscosity of the fluid to select the damping parameter $\tau = m/\gamma$ required for the thermostat employed in the LD simulations. For water, $\eta=8.90\times 10^{-4} \text{ Pa s}$ at 298 K. Using the definition of the friction coefficient, $\gamma=3 \pi \eta \sigma$,  gives $\tau_{_{water}}^{*} \approx 8\times10^{-4}$ in reduced units which corresponds to 0.8 ns in SI units, using the density of calcite 2710 kg/m$^3$, and nanoparticles of diameter 70 nm.

We used the thermostat for the Langevin equations of motion proposed by Bussi and Parrinello \cite{Bussi2007}, as implemented in LAMMPS \cite{Plimpton1995}. Due to the very steep and short range interactions, a very small timestep must be used to ensure accurate integration. To select the simulation time step, we monitored the conservation of effective energy \cite{Bussi2007}, defined as  $\widetilde{H} = E_{tot} - \Delta E_{tstat}$, where $E_{tot}$ is the total energy of the system and $\Delta E_{tstat}$ is the increment in the energy due to the thermostat. $\widetilde{H}$ was monitored for different values of timestep and damping parameters, for the inter-particle interaction corresponding to surfaces with roughness $\rho=1.5$ \AA of particles with $\sigma=70 \text{nm}$. We find good 
energy conservation for $\Delta t = 1 \times 10^{-5}$ (see Fig. 1 in the Supplementary Information), in a range of damping parameters 1-100 $\tau_{_{water}}$. 
Larger damping parameters correspond to lower viscosities, thus effectively increasing the efficiency of the simulation. For this reason, we set $\tau = 100\tau_{_{water}}$. While the change in the effective viscosity of the suspension modifies the dynamics ( not investigated in this work), it does not influence the final structure of the colloidal suspension.

\section{Results and discussion}

\subsection{Impact of roughness on nanoparticle-nanoparticle interactions}

Starting from the surface free energy profiles for two atomically flat calcium carbonate surfaces shown in Fig.~\ref{fig:epitaxy}, we calculated, using the Derjaguin approximation and Eq. (\ref{eq:potential}) the 
hydration contribution to the interaction potential as a function of the centre-to-centre distance between nanoparticles of size $\sigma=70$ nm, addressing recent experimental studies of CN suspensions %
\cite{Liberto2019}. 
Further, we use 
Eq. (\ref{eq:dlvo}) to 
calculate the van der Waals interactions and the electrostatic double layer repulsion between the spherical nanoparticles.
We show in Fig.~\ref{fig:schematic}D the resulting interaction potentials for different concentrations of 
Ca(OH)$_2$, and in Fig.~\ref{fig:schematic}D for different 
surface roughness and c=0 mM conditions. 

 The addition Ca(OH)$_2$ has been shown to increase the zeta potential as well as inducing the electrostatic screening of the solution, thereby tuning the DLVO interaction of the colloidal suspension. The results are represented in reduced units, namely $U^{*} = U/k_{B}T$ with $T = 298K$, and $r^{*} = r/\sigma$, where $\sigma$ is the nanoparticle diameter.

Relatively low 
surface roughnesses, $\rho=1.5$~\AA~
result in deep energy minima, $\sim$ 
$20-30k_{B}T$. This interactions are similar to those considered before in models of 
adhesive hard-spheres and patchy colloidal potentials 
\cite{Miller2004,Lu2008,Foffi2005,Bianchi2006,Zaccarelli2007}, which included 
short-ranged attractive wells of the order of 10's $k_B$T.
Based on these works, 
we expect that the effective interaction employed here 
should lead to irreversible and diffusion-limited cluster aggregation (DLCA)~\cite{Lu2008}, characterized by the formation 
of a gel phase at low particle packing fractions, likely in the interval $\phi=0.01-0.10$~\cite{Griffiths2017}. Surface roughnesses below $1.5$~\AA~ should also result in irreversible aggregation.

Fig. \ref{fig:schematic}D shows that the strongest attraction
is obtained in the case of pure calcite. As the concentration of Ca(OH)$_2$ 
increases to 30 mM, the interaction strength 
decreases, and for higher concentrations, 
$c=50$ mM, the interaction strength increases again. The effect of surface roughness is shown in 
Fig. \ref{fig:schematic}E. Increasing the surface roughness results in a shift to longer distances of the repulsive branch associated to the 
hydration forces, with a concomitant reduction of the interaction strength (given by the potential well depth). The interaction depends strongly on the roughness and for small roughnesses ($\rho>2.0$~\AA), 
it becomes 
purely repulsive, while for smooth surfaces ($\rho=0$), adhesive forces due to the solvent structure lead to stable suspensions. Our analysis therefore shows that 
the conditions required to find a stable 
CN suspension will depend both on the concentration of Ca(OH)$_2$ (the concomitant change of the pH) and the surface roughness.
Experimental studies indicated that at concentrations of the order 30 mM the suspensions are stable, as indicated by the minimization of the storage modulus \cite{Liberto2019}.

\subsection{Impact of roughness on the phase behaviour of colloidal suspensions}

The phase behavior of the suspensions was investigated by  Langevin dynamics computer simulations (see \textit{Simulation methods} section, and Supplementary Information for further details). Starting from  initial configurations with a random distribution of colloids, we generated molecular dynamics trajectories spanning simulation times of $t^{*}=10^3$, which correspond to 500 $\tau_{D}$, where $\tau_{D} = R^2/D$ and $R$ and $D$ are the radius and diffusion coefficient of the nanoparticle. Relevant experimental times 
can be estimated using the diffusion coefficient, $D$, the viscosity of water at 298 K and the radius of the nanoparticles used in the experiments\cite{Liberto2017}, 35~nm, giving $\tau_{D} = R^2/D = R^3 6 \pi \eta / k_B T \sim 0.2$~ms.

\begin{figure*}[t]
\centering
A \includegraphics[width=0.30\linewidth,valign=t]{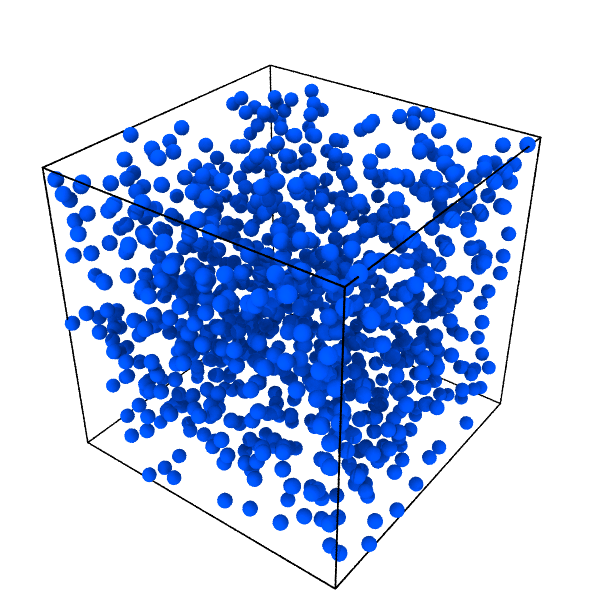}
B \includegraphics[width=0.30\linewidth,valign=t]{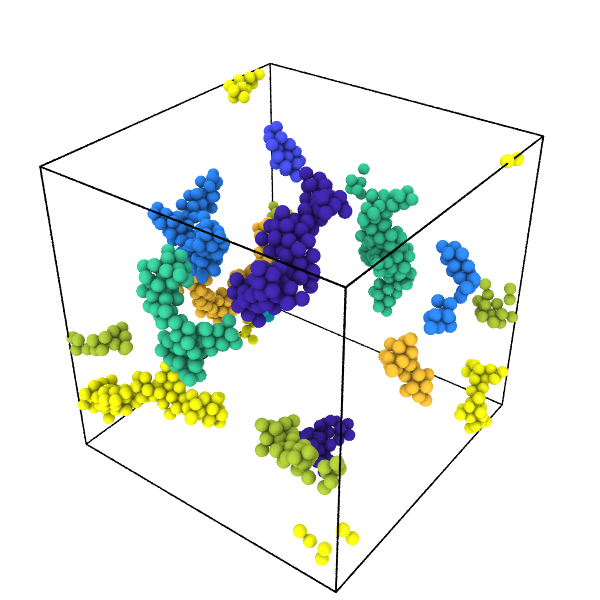}
C \includegraphics[width=0.30\linewidth,valign=t]{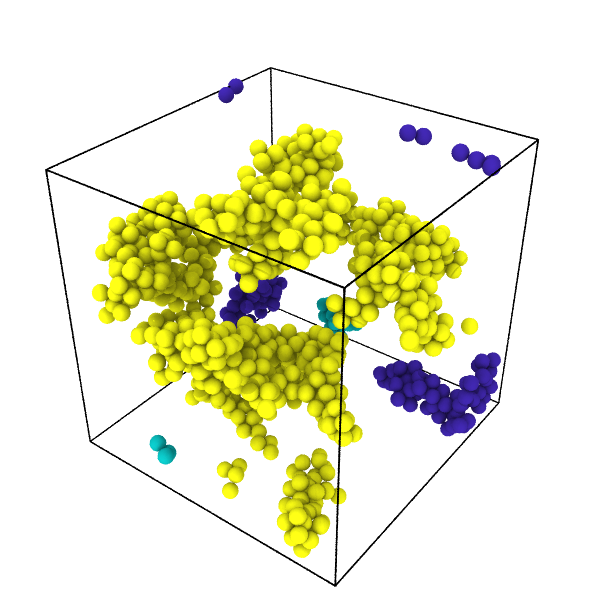} \\
D \includegraphics[width=0.30\linewidth,valign=t]{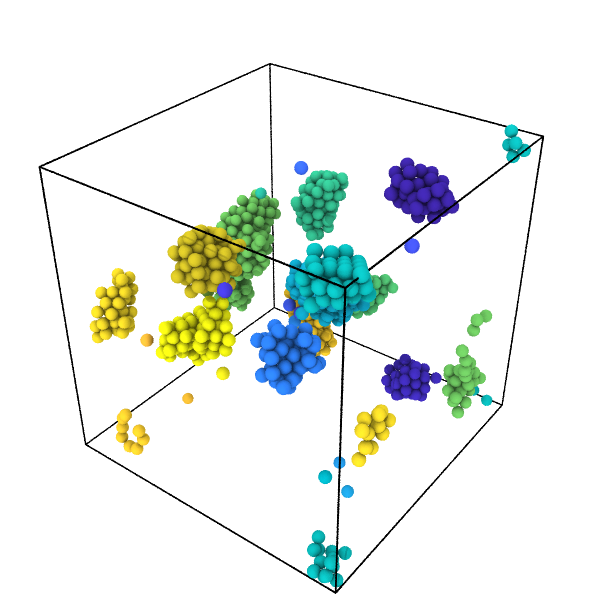}
E \includegraphics[width=0.50\linewidth,valign=t]{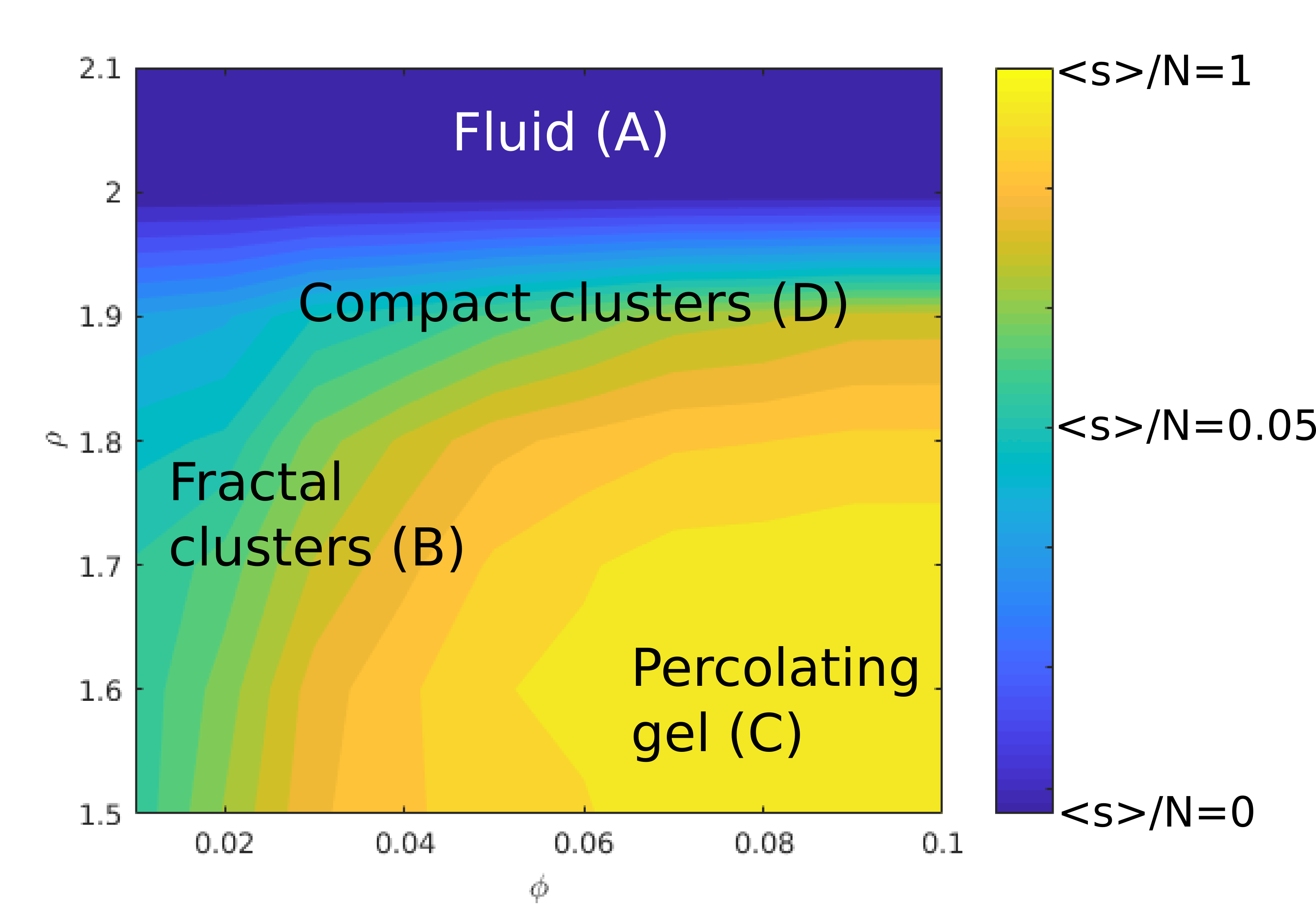}
\captionof{figure}{\footnotesize Characteristic structures of the suspension depending on volume fraction and surface roughness, for 30 mM Ca(OH)$_2$ concentration. A) Particles at packing fraction $\phi=0.05$, with surface roughness of $\rho=2.4$  \AA, showing a stable 
suspension. B) Particles at packing fraction $\phi=0.02$, with surface roughness of $\rho=1.5$  \AA, featuring different 
fractal clusters. Individual clusters have been identified using different colors. C) Particles at packing fraction $\phi=0.07$ and surface roughness of $\rho=1.5$  \AA, showing a percolating gel. In panels B-D, clusters have been identified with different 
colors. 
D) For a surface roughness of $\rho=1.8$  \AA, the adhesive energy is low enough such that the clusters reorganize into 
compact structures ($\phi=0.03$). E) Logarithm of the mean cluster size normalized by the number of particles, $\ln \left( \left\langle s \right\rangle/N \right)$, as a function of particle roughness, $\rho$, and particle volume fraction, $\phi$. 
}
\label{fig:suspensions}
\end{figure*}

Depending on the surface roughness, the CN formed 
a stable suspension (larger roughness $\rho>2.0$~\AA~, see Fig. \ref{fig:suspensions}A) or aggregated into fractal clusters (smaller roughness $\rho<2.0$~\AA~, see Fig. \ref{fig:suspensions}B and C). 
The interplay of the roughness, the adhesive and DLVO forces led to the formation of compact spherical aggregates (see Fig. \ref{fig:suspensions}D). The latter appear at high roughness in a wide range of volume fractions $\phi$. Intermediate roughnesses favour the formation of fractal and percolating clusters (see Fig. 3C).

To quantify the nanostructure of the nanoparticle suspensions we computed the mean cluster size,  employing a distance criterion to construct the clusters. Two nanoparticles $i$ and $j$ were assigned to the same cluster if their distance $d_{ij}<1.02 \sigma$. The attractive minima are contained within this distance, as seen in Fig. 2-D and 2-E, and thus corresponds to the characteristic distance between aggregated nanoparticles. The mean cluster size distribution, $\left< s \right>$, was computed using the equation~\cite{Sevick1988,Sciortino2005}:

\begin{equation}
 \left\langle s \right\rangle = \dfrac{\sum\limits_{s=1}^{s_{max}}s^{2}P(s)}{\sum\limits_{s=1}^{s_{max}}sP(s)}
 \label{eq:meancluster}
\end{equation}

where $s$ is the cluster size, $P(s)$ is the cluster size probability distribution and the sums run over all clusters from size 1 up to the maximum size, $s_{max}$, observed in the simulations.

\subsection{Cluster structure and Fractal dimension}
The fractal dimension (FD) of the cluster formed in nanoparticle suspension is commonly used as a
``fingerprint" of the structure of the suspension. The FD can be inferred from the analysis of
experimental data obtained with light scattering experiments~\cite{Burns2002}, and theoretically from
the analysis of the simulation trajectories.
We computed the Hausdorff fractal dimension \cite{Mandelbrot1989} 
using the box counting algorithm \cite{Gagnepain1986}.
The simulation box was divided into an integer number of cubic cells, $n$, each cell with 
length $l = L/n$. The number of cells occupied by colloids, $N_{f}$ was then calculated by monitoring the coordinates of the colloids and thus the cell they belong to, and the calculation was repeated for boxes of different lengths, $N_{f}(l)$ 
(see SI for an example of this calculation). The fractal dimension, $d_{f}$, is then obtained from  \cite{Griffiths2017}:

\begin{equation}
 d_{f} = \dfrac{\ln N_{f}(l) }{\ln (1/l)}
\end{equation}

The  fractal dimension of the suspension was calculated using clusters containing more than 20 colloids, with a similar expression to Eq. (\ref{eq:meancluster}):

\begin{equation}
 \left\langle d_{f} \right\rangle = \dfrac{\sum\limits_{s=20}^{s_{max}} d_{f} s P(s)}{\sum\limits_{s=20}^{s_{max}}sP(s)}
\end{equation}

Following the work by Griffiths \textit{et. al}, we  calculate the fractal dimension at two scales. The first scale corresponds to box sizes with a size similar to the characteristic size of the particle ($l=[1 \sigma, 5\sigma]$), hence  
probing the local structure of the cluster (\textit{local fractal dimension}), and 
characterizing the degree of compactness of individual clusters. A second scale corresponds to box counting cell sizes in the range $l=[5 \sigma, L]$, where $L$ is the size of the simulation box. This length scale quantifies the amount of volume fraction 
occupied by each cluster, and therefore it is related with the percolation of the clusters (\textit{global fractal dimension}). See SI for more information with respect to \textit{local} and \textit{global} fractal dimensions.

\begin{figure*}[t]
\centering
\includegraphics[width=0.8\linewidth]{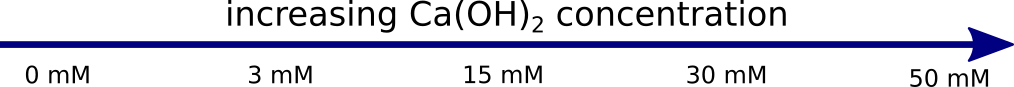}
\\
\rotatebox{90}{$\qquad ln(\langle s \rangle/N)$}
\includegraphics[height=0.13\linewidth,valign=b]{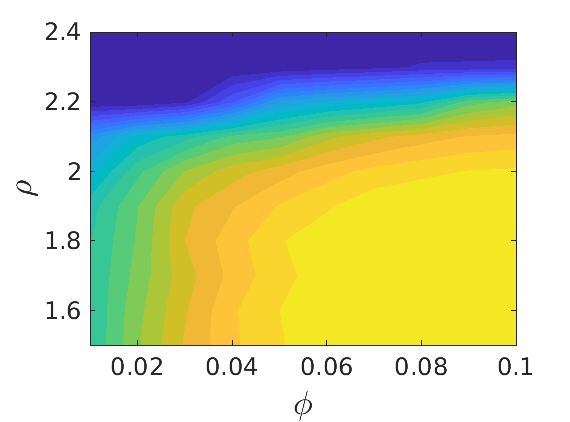}
\includegraphics[height=0.13\linewidth,valign=b]{figures/meansize_2.png}
\includegraphics[height=0.13\linewidth,valign=b]{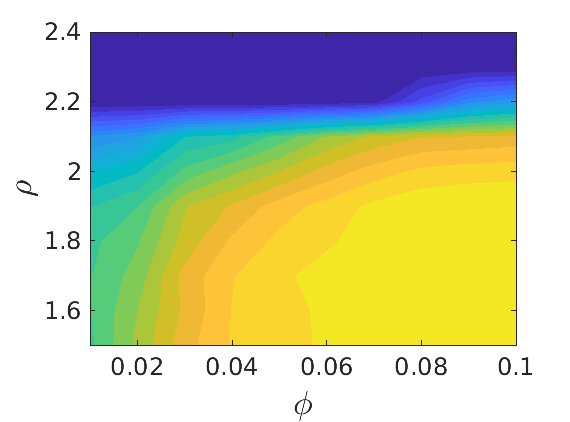}
\includegraphics[height=0.13\linewidth,valign=b]{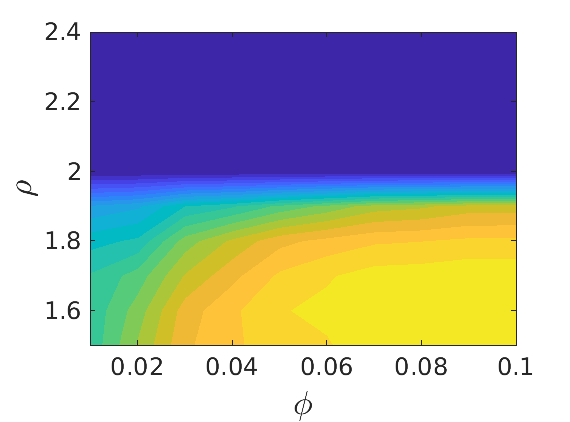}
\includegraphics[height=0.13\linewidth,valign=b]{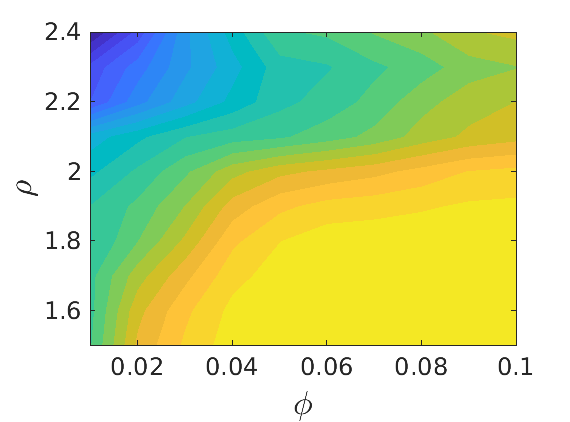}
\includegraphics[height=0.13\linewidth,valign=b]{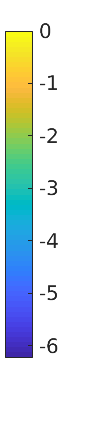}
\\
\rotatebox{90}{$\qquad \qquad d_{f}^{g}$}
\includegraphics[height=0.13\linewidth,valign=b]{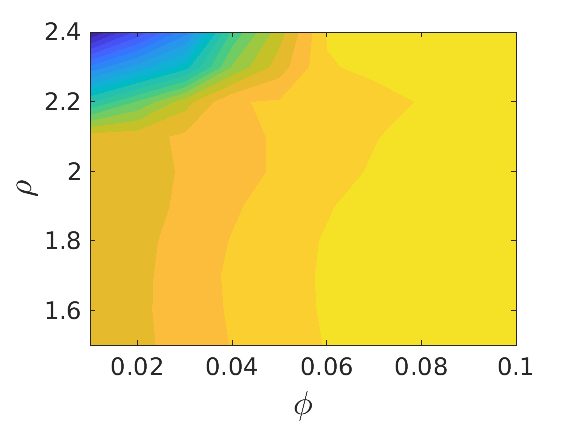}
\includegraphics[height=0.13\linewidth,valign=b]{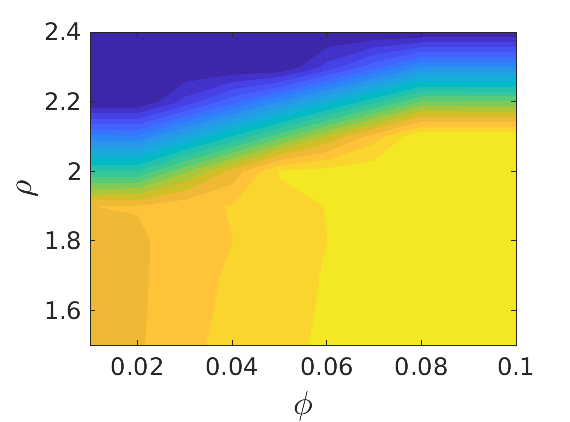}
\includegraphics[height=0.13\linewidth,valign=b]{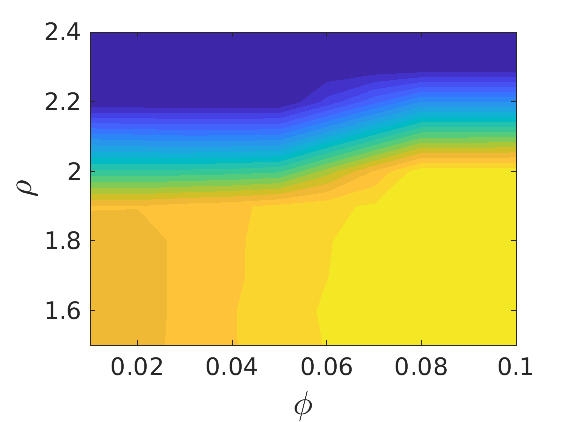}
\includegraphics[height=0.13\linewidth,valign=b]{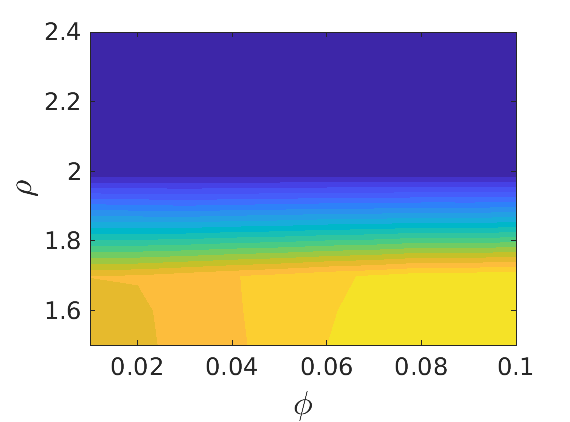}
\includegraphics[height=0.13\linewidth,valign=b]{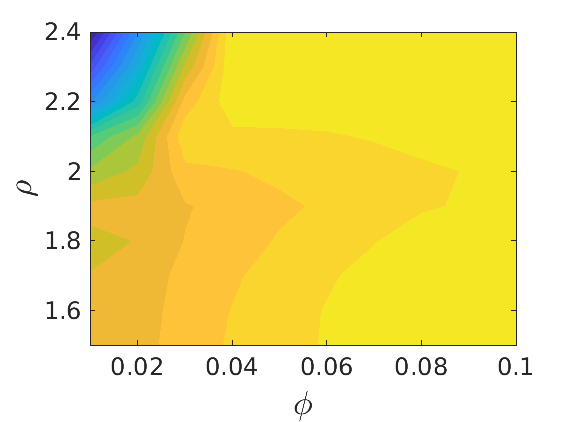}
\includegraphics[height=0.13\linewidth,valign=b]{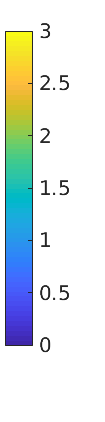}
\\
\rotatebox{90}{$\qquad \qquad d_{f}^{l}$}
\includegraphics[height=0.13\linewidth,valign=b]{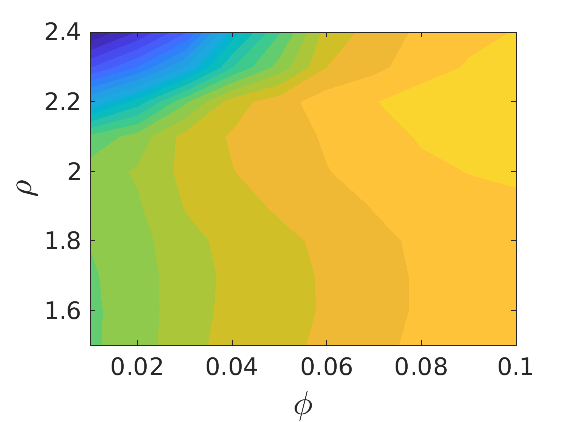}
\includegraphics[height=0.13\linewidth,valign=b]{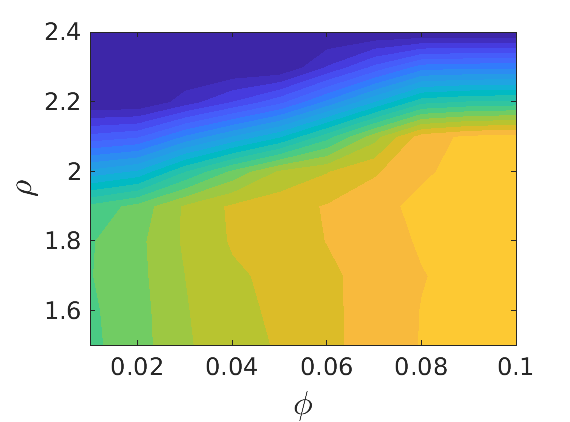}
\includegraphics[height=0.13\linewidth,valign=b]{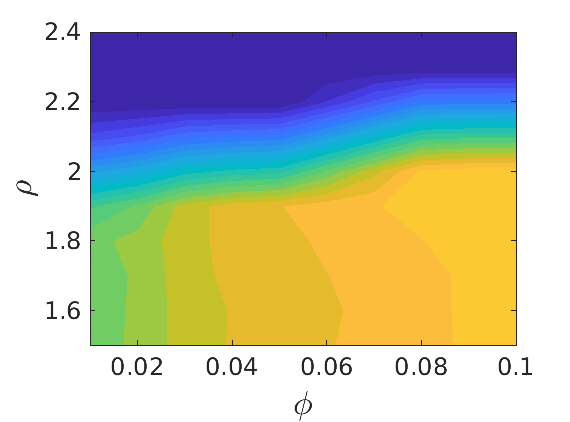}
\includegraphics[height=0.13\linewidth,valign=b]{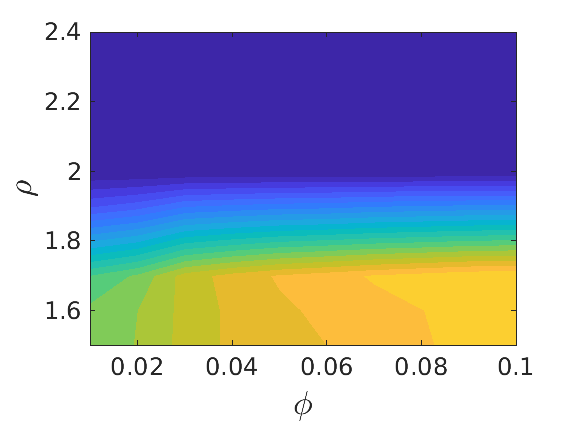}
\includegraphics[height=0.13\linewidth,valign=b]{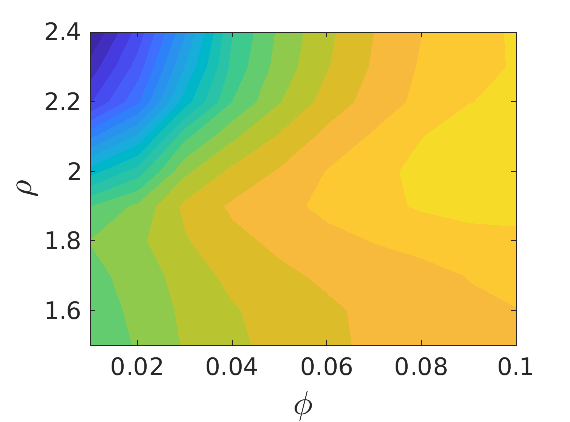}
\includegraphics[height=0.13\linewidth,valign=b]{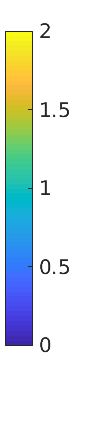}
\captionof{figure}{\footnotesize Logarithm of the mean cluster size normalized by the number of particles (top).  Global (middle) and local (bottom) panels represent fractal dimensions as a function of the packing fraction, $\phi$, and surface roughness, $\rho$, for different concentrations of 
Ca(OH)$_2$. For $c=30$ mM, the surface roughness threshold for a stable suspension reaches the smallest value . 
the mean cluster size and the global fractal dimension feature consistently the lowest values too. The degree of compaction, characterised by the local fractal dimension, shows a maximum around $\rho=2.0$~\AA~at $c=0$ mM and $c=50$ mM, owing to the decrease in the interaction strength, which favours the cluster 
reorganisation into more compact structures. The color maps were constructed using a running average over 3 points.}
\label{fig:fractal}
\end{figure*}

We show in Figure~4 a colour map that illustrates the range of mean cluster size and fractal dimensions in the roughness/packing fraction plane.
The mean cluster size as well as global the fractal dimension show that the cluster fill the space more efficiently at high volume fractions 
(large fractal dimension), and form 
percolating structures with a percolation threshold between $\phi= 0.04-0.05$, as evidenced by an increase in the mean cluster size $\langle s \rangle$ converging to $\sim N$, the number of particles in the simulation cell.
For volume fractions $\phi > 0.05$ and roughness values $\sim 2.0$ \AA, the structures transition from a percolating gel to a fluid phase. This region of the phase diagram is characterized by an increase in the local fractal dimension,  $d_{f} \sim 1.8$. The increase in the fractal dimension reflects an increase in the degree of compaction of the clusters, which arises from 
weaker attractive interaction. The reduction in the interactions allows the clusters to rearrange themselves into more compact structures, specifically crystalline structures, as shown in Fig. \ref{fig:suspensions}-D. Aggregation into local crystalline structures as opposed to the formation of percolating gels, has also been observed by Griffiths and coworkers \cite{Griffiths2017} using the Morse pair potential with different interaction strengths. Those authors reported local fractal dimensions similar to the ones we find here. Similar fractal dimensions have also been reported in experimental studies of gold colloidal aggregates formed via irreversible kinetic aggregation \cite{Weitz1984}. This supports our view that the percolating phases form following diffusion-limited cluster aggregation.

Our simulations show that at fixed roughness, the addition of the Ca(OH)$_2$ and therefore the increase in the solution pH, results in non-monotonic changes of the fractal dimension with volume fraction. For instance we observe a minimum in the fractal dimension at 30~mM and $\rho=1.7$, which reflects the formation of a stable suspension, while at this roughness results most of the suspensions at other concentrations feature
adhesive behaviour.

\subsection{Discussion and Conclusions}
In summary, we have investigated the stability of suspensions of \blue{models of calcium carbonate nanoparticles 
using effective potentials that incorporate hydration and double layer forces. We have introduced interaction potentials between calcium carbonate surfaces by combining the adhesion forces arising from water hydration layers, with DLVO interactions inferred from experimental studies. We propose that the adhesive interaction, predicted recently using fully atomistic simulations of calcium carbonate surfaces, must be included to model the phase behavior of the nanoparticle suspensions. We find that small changes in surface roughness can inhinite adhesion. This behavior can be rationalized by considering the narrow range of the hydration interactions acting between calcite nanoparticles at short distances. We used Langevin Dynamics simulations to investigate the phase diagram of the suspensions and to calculate the fractal dimension of the nanoparticle clusters, as a function of nanoparticle volume fraction, nanoparticle roughness and  concentration of Ca(OH)$_2$, or pH of the solution.}

Our results show that nanoparticle surface roughness has a large impact on the inter-colloidal interactions and on the stability of the suspensions. \blue{Our model predicts aggregation and gel formation (as reported in experiments of calcite nanoparticles \cite{Liberto2019}) when the surface  roughness is small ($\rho<2.0$~\AA~)}. The fractal dimension of the clusters obtained in this way is comparable to the diffusion-limited cluster aggregation fractal dimension of $\sim 1.75$ \cite{Burns2002}, where particles aggregate when undergoing a random walk, without further reorganization of the structure upon joining the nanoparticle cluster. Increasing the surface roughness leads to a shift of the effective repulsion to longer inter-particle distances and to the stabilization of the suspension.
We have observed a synergistic effect between nanoparticle roughness and DLVO forces (mediated by changes in the concentration of Ca(OH)$_2$). At conditions corresponding to high base concentrations, $50$~mM of Ca(OH)$_2$, attractive interactions may be important, even for relatively large roughness, leading to the aggregation of the suspensions. At intermediate concentrations, 30 mM, the suspensions are stable, even for relatively low surface roughness, and the suspensions features a minimum in the roughness required for gel formation. Our model predicts that at lower concentrations, $<30$mM, gel phases are formed for small roughness. \blue{High pH conditions and high ionic concentration (high Ca(OH)$_2$) may disrupt the particle morphology inducing an increase in the roughness. According to our model such effects would eliminate the adhesive minimum, making aggregation less likely.} 
The lack of gel phases at 30 mM concentration is consistent with the experimental measurements of calcium carbonate nanoparticle suspensions, which reported a minimum in the storage modulus at this concentration \cite{Liberto2019}, signalling the formation of a fluid phase. Our model indicates that this phase can be formed because the nanoparticles are sufficiently rough (according to our model $\rho > 1.5$ ~\AA), since roughness below this value result in the formation a gel phase. This result should motivate additional experimental analyses to resolve the surface structure of calcite nanoparticles.

Overall, our result highlight the importance of the surface structure
of nanoparticles on the phase behavior of suspensions. \blue{We provide theoretical predictions of the relationship between the stability of the suspension and the inter-colloidal energy profiles, with particular emphasis on the modification of the adhesive minimum}.  
\blue{We have illustrated the interplay between roughness and adhesive effects in a models} of calcite nanoparticle suspensions, which are relevant in
industrial applications, building materials and biomineralization processes.
Our results indicate that small changes in the surface roughness \blue{of the nanoparticles} (of the order of calcite atomic steps) might modify significantly the interparticle forces.
Indeed, we \blue{demonstrate} that attraction or repulsion between the surfaces can be observed with the same underlying interactions, when the roughness of the surface is taken into account. We expect that our work will serve to rationalize existing contrasting observations, as well as to connect the phase behavior of suspensions to the surface topography and volume fraction of the suspension. The phase diagram reported in this work (Figure 3-E) provides a route to establish this connection. \blue{While, we have focused on calcite, by considering a combination of interactions obtained from atomistic simulations and experimental studies, the theoretical approach discussed here to predict the relationship between the stability of the suspension and the inter-colloidal interactions, could be extended to other colloidal suspensions}. 

\paragraph*{Acknowledgements}
We acknowledge the EPSRC-UK (Grant No. EP/J003859/1), the EU NanoHeal ITN project grant agreement No. 642976 for financial support. We thank the Imperial College High Performance Computing Service for providing computational resources. We thank Dag Dysthe, Marie Le Merrer, Teresa Liberto and Catherine Barentin for illuminating discussions.

\section*{Conflicts of interest}
There are no conflicts to declare.

\section*{Supplementary information}

\subsection*{Probability distribution of surface heights}

The continuous probability distribution for the surface heights follows a Gaussian distribution, given by:

\begin{equation}
 f_{h} = \dfrac{e^{-(r-h)^{2}/(2 \rho^{2})}}{\rho \sqrt{2 \pi}}
 \label{eq:roughnesssup}
\end{equation}

\noindent where $r$ is the reference surface, $h$ is the deviation with respect to $r$, and $\rho$ is the surface roughness ($\rho=0$ corresponds to a flat surface). 

\begin{figure}[H]
\centering
\includegraphics[width=0.8\linewidth]{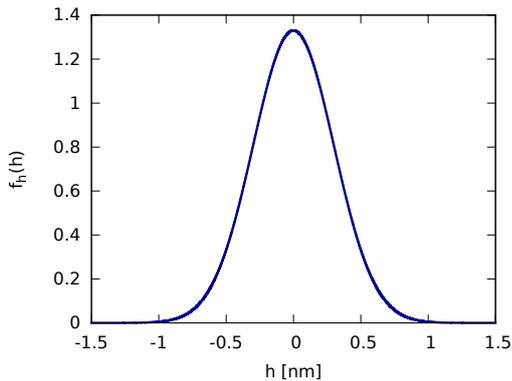}
\captionof{figure}{\footnotesize Surface height distributions, with $\rho=0.3$ nm.}
\label{fig:probability}
\end{figure}

Fig. \ref{fig:probability} shows the surface height distribution, with roughness $\rho=0.3$ nm.

\subsection*{Effective interparticle interactions}

\begin{figure}[H]
\centering
\includegraphics[width=0.8\linewidth]{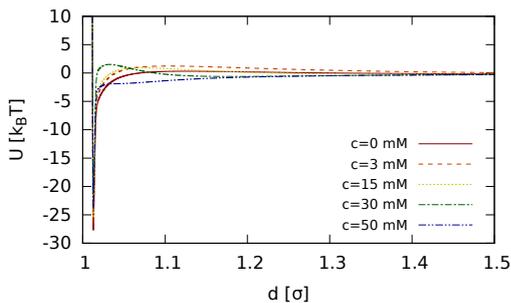}
\captionof{figure}{\footnotesize Inter-particle interactions as a function of Ca(OH)$_2$, in the range of inter-particle distance $1.0\sigma-1.5\sigma$.}
\label{fig:longrange}
\end{figure}

Fig \ref{fig:longrange} shows the effective inter-particle interactions in a longer range than that considered in Fig. 2D in the main document. 

\subsubsection*{Particle size dependence of the effective interactions}

\begin{figure}[H]
\centering
\includegraphics[width=0.8\linewidth]{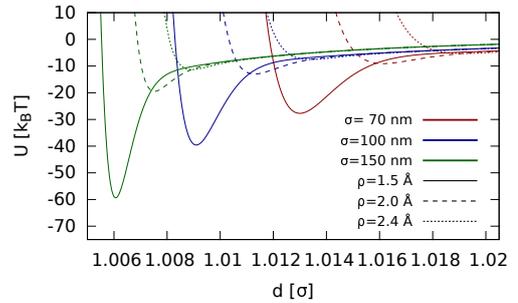}
\captionof{figure}{\footnotesize Inter-particle interactions as a function of particle size $\sigma$ and surface roughness $\rho$.}
\label{fig:sizedependence}
\end{figure}

\blue{The dependence of the effective interactions on the particle size is shown in Fig. \ref{fig:sizedependence}. The increase of nanoparticle size results in an increase of the potential well depth, which has the consequence of increasing the minimum surface roughness required to stabilize the nanoparticle suspension. Larger particles featuring the same surface roughness, for instance $\rho=1.5$~\AA~ should therefore tend to aggregate with increasing adhesive energies.}

\subsection*{Langevin dynamics}

\begin{figure}[]
\centering
\includegraphics[width=0.8\linewidth]{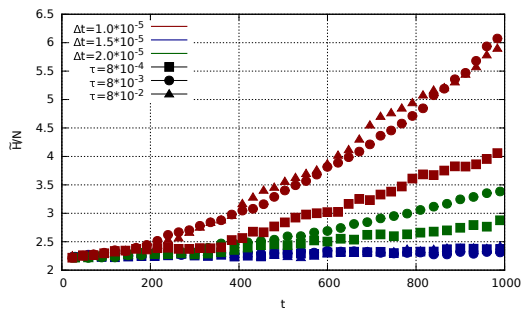}
\captionof{figure}{\footnotesize Time evolution of the effective energy per particle, as a function of the timestep and the damping parameter, for the smooth potential of particles with $\sigma=70 \text{nm}$.}
\label{fig:effective_timestep}
\end{figure}

To account for the stochastic collision of the solvent with the nanoparticles, and therefore Brownian motion, we used Langevin dynamics~\cite{vanKampen1981}. In this method, a friction force and noise term are added to the equations of motion:

\begin{eqnarray}
 \dfrac{d {x}(t)}{dt} & = & {v}(t) \\
 m \dfrac{d {v}(t)}{dt} & = & -\zeta {v}(t) -\dfrac{d {U}}{dt} + {F}(t)
  \label{eq:langevin}
\end{eqnarray}

\noindent where ${x}$ and ${v}$ are the position and velocity of a nanoparticle, $m$ is its mass, $\zeta$ is the friction coefficient, ${U}$ is the interaction energy between particles, and ${F}(t)$ is the stochastic noise term. This noise term fulfills $\left\langle {F}(t) \right \rangle = 0$ and $\left\langle {F}(t_{1}){F}(t_{2}) \right \rangle = \Gamma (t_{1}-t_{2})$, where $\Gamma=2 \zeta k_{B} T$ quantifies the strength of the stochastic noise \cite{vanKampen1981}. We used the LAMMPS \cite{Plimpton1995} implementation of the thermostat proposed by Bussi and coworkers \cite{Bussi2007}. In order to verify the accuracy of the integration, we used the effective energy conservation defined as \cite{Bussi2007}:

\begin{equation}
 \widetilde{H} = E_{tot} - \Delta E_{tstat}
\end{equation}

\noindent where $E_{tot}$ is the total energy of the system, and $\Delta E_{tstat}$ is the increment in energy due to the thermostat. We monitored the effective energy conservation for different values of timestep and damping parameters (see Fig. \ref{fig:effective_timestep}), for the interaction potential of particles with $\sigma=70 \text{nm}$, surface roughness $\rho=1.5$~\AA~, and CaOH$_2$ concentration $c=0$ mM, which corresponds to the steepest and strongest attractive potential. We find effective energy conservation for $\Delta t = 1.0 \times 10^{-5}$, in a range of damping parameters 1-100 $\tau_{_{water}}$, where $\tau_{_{water}} \approx 8\times10^{-4}$ in reduced units which corresponds to 0.8 ns in SI units, using the density of calcite 2710 kg/m$^3$, and the nanoparticles of diameter 70 nm.

\subsection*{Calculation of fractal dimension}

We present an example of the box counting algorithm, for the percolating gel shown in Fig. 3C in the main document. The system is divided in $N_{b} = 1,2,3 \dots$ segments on each dimension, and the number of filled boxes is counted. The log-scale plot of the number of filled boxes $N_{f}$ as a function of $L/l$ gives the fractal dimension, which is probed on the ranges of $l=[1 \sigma, 5\sigma]$ for the local fractal dimension, and $l=[5 \sigma, L]$ for the global fractal dimension. The variation in the slope at $\approx 5 \sigma$ corresponds to the transition between the global structure of the cluster (\textit{i.e.} its space-filling characteristics) and the local structure of the cluster.

\begin{figure}
\centering
\includegraphics[width=0.8\linewidth]{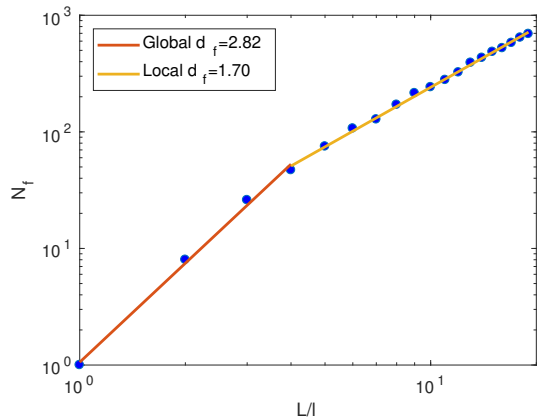}
\captionof{figure}{\footnotesize Number of filled boxes as a function of the inverse of the box length normalized by the simulation cell size $L$, for the percolating cluster at $\phi=0.07$, $\rho=1.5$~\AA~ and CaOH$_2$ concentration $c=30$ mM (shown in the main document as the structure in Fig. 3C. The global fractal dimension is calculated with the slope in the range $l=[5 \sigma, L]$ and the local fractal dimension is calculated with the slope in the range $l=[1 \sigma, 5\sigma]$.}
\label{fig:fractal_example}
\end{figure}

\balance

\bibliographystyle{rsc}
\bibliography{roughness.bib}

\balance

\end{document}